\documentclass[a4paper]{jpconf}

\usepackage[english]{babel}
\usepackage{wrapfig}
\usepackage{custom}

\begin{document}
\title{Relativity and Quantum Theory:\\
Under the Spell of Today's Paradigms}

\author{Stefan Weigert}

\address{Department of Mathematics, University of York, York, United Kingdom}
\ead{stefan.weigert@york.ac.uk}

\begin{abstract}
Thomas S. Kuhn interprets the development of the (natural) sciences
as a specific dynamical process. Periods of piecemeal growth of knowledge
based on widely accepted paradigms are interrupted by bursts of revolutionary
changes which lead to new paradigms incommensurate with the earlier
ones. This process is briefly illustrated by recalling the changes
to classical physics brought about by Einstein's theory of relativity on the one hand, and by quantum theory on the other. Both
theories represent fundamental paradigms of contemporary physics.
They appear unshakable To the working physicist but according to
Kuhn their paradigmatic status is of a temporary nature only. Does Kuhn's framework help us to identify potential
future revolutions?
\end{abstract}

\section{Introduction}

By its very nature, history is descriptive and comes after the fact.
Nevertheless, there is a wide variety of both methods applied and
goals declared by historians. The 19th century saw speculative theories
which postulated forces driving mankind towards a specific future
state of society, or they assumed some necessary ``dialecti'' evolution. However, the idea of ``laws'' acting throughout history
was met with considerable skepticism in the 20th century, with philosophers of history cutting back such large-scale interpretations.
Instead, the focus turned towards fairly representing past events, avoiding to judge them based on a privileged
-- later -- position and not distorting them in order to support some preconceived idea. These views clearly leave no room for the idea that paste events could reliably determine future events, in spite of being strongly intertwined.

Thomas S. Kuhn's views run counter to this trend, albeit in the narrow
setting of the history of science where it might be possible to say
more \cite{KuhnRevolutions}. Kuhn suggests that scientific theories -- physics
in particular -- evolve according to a ``binary'' pattern consisting
of longer periods of \emph{normal science} time interrupted by shorter
burst of revolutionary \emph{paradigm shifts}, induced by \emph{anomalies} which
create \emph{crises}. This conception represents a return to 19th-century
ideas in the sense that there exists a \emph{law} -- or at least a
definite \emph{pattern} of ``scientific revolutions'' -- which
is supposed to govern the way in which sciences develop over time. This approach
is at odds with the idea that scientific activities would be purely
additive or cumulative, a view shared by many earlier historians
of science. However, Kuhn does not suggest that the proposed structured
progression would lead to a specific final state. Being ``goal-less'',
the evolution rather has Darwinian traits. 

For the sake of the argument, let us assume that Kuhn's structure
of scientific evolution correctly describes the relation of theories
following each other. The presence of law-like development -- even
if as rudimentary as a sequence of normal and revolutionary periods -- raises
the question whether the model possesses any ``predictive'' power.
The interesting point here is to ask whether the model may help scientists
to judge the status of a currently valid paradigm which defines their
world view. In other words, is Kuhn's model limited to retrospective
descriptions only -- once paradigms have lost their power -- or may
it be used to identify potential future revolutions from within a
given paradigmatic framework?

To answer this question, we will first use Kuhnian terminology to
describe the transition from classical Newtonian physics to today's
theory shaped by Einstein's theory of relativity and by quantum theory.
In the process, fundamental concepts such as simultaneity and particle trajectories were given
new meanings incompatible with those of the pre-revolutionary framework. The
resulting post-revolutionary framework, based on quantum theory and
relativity, defines the currently accepted paradigm of physics. We will then be in a position to ask whether
Kuhn's model allows us to transcend the paradigms of today's physics.

In Sec.~2, we briefly describe the key notions of Kuhn's dynamical
model of the natural sciences: \emph{paradigms} which underlie normal
science,\emph{ anomalies} within a given set of paradigms, and scientific
\emph{revolutions} triggered by the growing weight of inconsistencies.
In Sec.~3, these concepts are illustrated by developments in physics
which happened at the beginning of the 20th century when these counter-intuitive theories
toppled the long-standing paradigm
of Newtonian classical mechanics. These scientific revolutions of
the past set the scene for potential future revolutions which would
have to fundamentally alter relativity and quantum theory, or replace
them. Sec.~4 discusses the current status of these theories while Sec.~5 addresses the 
question to what extent Kuhn's model may help us to prepare the next paradigm
shift.

\section{Paradigms, anomalies and revolutions}

A survey of the technologies which reduced the necessity of manual labour, simplified communication or
facilitated easier transport, for example, may suggest that they 
evolve in a cumulative way. Improvements are often based on refining
existing methods or devices. It seems natural to assume an equally
linear and additive development of the sciences
which underpin the technological changes. According to Kuhn, a closer
reading of the history of science reveals that this view is inappropriate -- it does not 
do justice to the dynamics of fundamental sciences such as
physics. 

A simplified version of Kuhn's main idea 
splits scientific activity into two distinct phases which inevitably
alternate: \emph{normal} and \emph{revolutionary }periods. Widely
accepted \emph{paradigms }\cite[Sec.~V]{KuhnRevolutions} define
a framework within which main stream research takes place -- normally
over an extended period -- by providing concepts and theoretical tools
which\emph{ }a community of researchers shares. A consistent ``world
view'' determines both relevant problems worthy of investigation
and the methods available to study them. Kuhn refers to this activity
as ``puzzle-solving'' \cite[Sec.~IV]{KuhnRevolutions}. In other
words, the consequences of an unquestioned framework are being unfolded
in detail, without worrying too much about its foundations. 

According to Kuhn, the explanatory power of a framework tends to reach
its limits. After a period of normal scientific activity, which sees
its successful application and expansion, some experimental data
may be found to disagree with the predictions of the framework, or
internal inconsistencies are discovered. Such \emph{anomalies} \cite[Sec.~VI]{KuhnRevolutions}
are often ignored by the majority of researchers as long as they do not directly affect their
field. 

If more anomalies are being found and cannot be resolved within the
dominant paradigm, they lead to an increasing number of researchers
questioning its validity. A scientific \emph{revolution }\cite[Secs.~IX, X]{KuhnRevolutions}
has occurred once an alternative framework has been put in place which
will have similar explanatory power as the earlier one while eliminating the
anomalies. Typically, the new paradigm is \emph{incommensurate} with
the old one in the sense that it does not simply extend or adapt the
known concepts; a disruptive conceptual break takes is necessary to
provide new foundations.

Kuhn's binary scenario -- periods of normal activity alternate with revolutions triggered by anomalies which
necessitate paradigms incommensurate with the earlier ones -- has attracted considerable interest. It has been debated extensively  
 by historians and philosophers of science as well as sociologist (see e.g. \cite{KuhnReception}). 

For our purpose we will assume that Kuhn's view captures essential
aspects of how theories such as physics evolve. We begin by illustrating
the model and its central notions by briefly describing
two developments within physics which happened early on in the 20th
century: the emergence of the theory of special relativity and of
quantum theory. Only then will we be able to address the main question.

\section{Past Revolutions}

\subsection{The paradigm: Newtonian physics}

Newton's \emph{Principia} \cite{NewtonPrincipia} established the foundations of
classical mechanics in 1689.  The theory relies on primitives such as time and reference
frames which were not challenged for more than two centuries. The behaviour of material bodies is governed by dynamical laws which require a definition of \emph{time}. In the \emph{scholium}, an early part of his \emph{Principia}, Newton introduces time in an axiomatic way,
\begin{quote}
\emph{Absolute, true, and mathematical time, of itself, and from its own nature, flows equably without relation to anything external} {[}$\ldots${]}. \cite[p~6]{NewtonPrincipia}.
\end{quote}
This definition guarantees an unambiguous ordering of events by means
of the universal parameter time. A sunrise observed at the location of Newton's
apple tree, say, occurs at a specific moment of time. Any other event
will either happen earlier, later or at the same time -- the ordering of events
is unambiguous when referring to absolute time. 

Newton's laws predict the future motion of an object
under the influence of known external forces with certainty and 
arbitrary precision given its exact \emph{position} and its\emph{velocity}
at present. Thus, at any given moment of time, a small particle, say, is assumed to occupy a specific position in space
and to move into a certain direction with well-defined speed. These data define its \emph{state}. In turn, the state can be determined unambiguously by suitable position and velocity measurements. It will not change since the interaction with the particle can, in principle, be made arbitrarily weak.

The motion of celestial -- and all other -- bodies takes place relative
to the \emph{aether}, an unobservable substance permeating the
entire universe. The aether was assumed to exist to provide a medium for light
to propagate through vacuum, in analogy to waves travelling on the
surface of water, or air carrying sound waves. 

\subsection{Revolution 1: Theory of relativity}

\label{subsec:Revolution relativity}

Experiments to determine the motion of the earth relative to the aether were conducted throughout
the second half of the 19th century, with ever increasing precision. In 1887, Michelson and Morley devised a method to to test whether the speed of light would depend on its direction of travel relative
to the aether \cite{MichelsonMorley}. Based on interfering light
beams, their observations were sufficiently precise to rule out such
a dependence. The negative result represented a major inconsistency
within the framework of classical mechanics. To account for the result,
contrived-looking assumptions about the propagation of light had to
be made.

Paradigm change came about in 1905 when Einstein scrutinized the concept
of time used in Newtonian mechanics,
\begin{quote}
\emph{We must take into account that all our judgments in which time plays a part are always judgments of} simultaneous events. \cite[p~893]{EinsteinSRT1905} 
\end{quote}

Reference to the absolute time introduced by Newton turns out to be
insufficient to define simultaneous events: it is not obvious 
how to ensure that clocks in different locations actually display
the same time. What exactly does it mean to say that events do happen at the same time?
Einstein addresses this question by proposing an \emph{operational} 
approach to synchronize clocks, i.e. by spelling out
a physical procedure for people to synchronize their clocks. The observers
are assumed to operate within \emph{inertial reference frames} which
move relative to each other along straight lines and at constant speed.
Einstein's method to synchronize clocks is based on a new postulate compatible
with the findings of Michelson and Morley: the speed of light is constant
in all inertial references frames. This assumption establishes the
(satisfying) physical equivalence of all such frames but implies, somewhat counter-intuitively, that
two events happening at the same time for one observer are not necessarily
simultaneous for another one.

Eliminating simultaneity based on absolute time is a ``revolutionary''
conceptual step as it does away with one of the fundamental notions
on which classical mechanics is built. Newton's conception of time is replaced by a new notion, ``incommensurate''
with absolute time. The change was triggered by the inexplicable, ``anomalous'' outcome of
an experiment which the new paradigm, the theory of relativity, was able to predict correctly. 

\subsection{Revolution 2: Quantum Theory}

The first quarter of the 19th century saw another paradigm shift.
Microscopic objects were found to behave in ways which could not be
explained by Newtonian mechanics. Atoms were found to be aggregates
of particles with opposite electric charge. Classical models imply that they must attract each
other, rendering the atoms unstable, an obvious contradiction with the
permanent material structures we live in. Experiments suggested that atoms are emitting
\begin{wrapfigure}{o}{0.5\textwidth}%
\begin{centering}
\includegraphics[angle=90,origin=c,width=0.4\columnwidth]{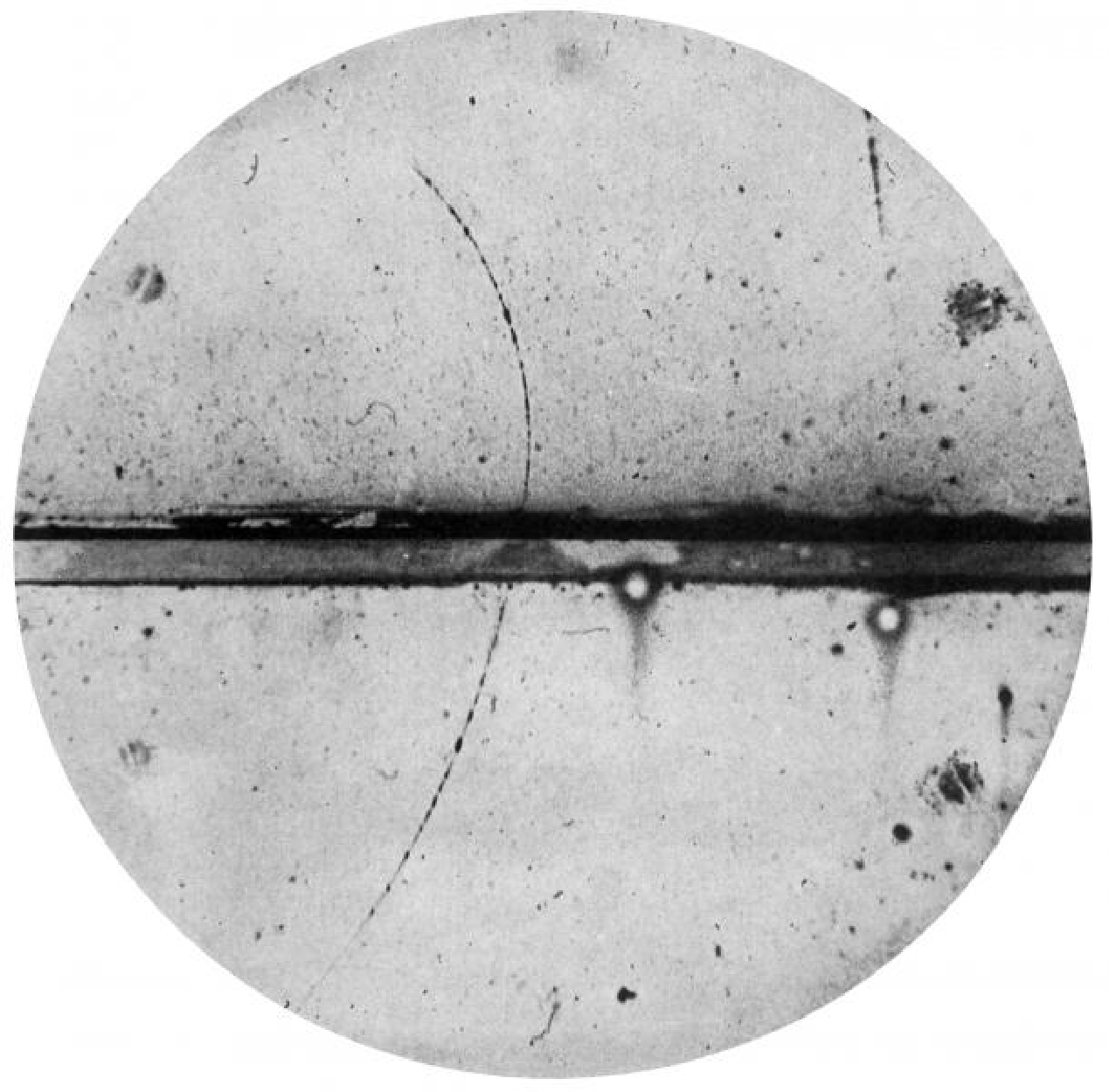}
\par\end{centering}
\caption{\label{fig:Cloud-chamber-record} The first record of a positron in the presence of a magnetic field; cloud chamber photograph from 1933 \cite{cloudchambertrack}.}
\end{wrapfigure}%
radiation of specific frequencies only, a property impossible to derive from classical
mechanics. Many other anomalous properties of microscopic objects could not be explained using the age-old paradigm.

In 1927, Heisenberg takes stock of these developments and analyses the observed anomalies 
from a conceptual point of view \cite{HeisenbergSketch}. In particular, he contemplates
experiments with so-called cloud chambers. These devices, dating from the early 1900s, were essential to understand
the properties of small particles invisible to the
naked eye, such as electrons (cf. Fig.~\ref{fig:Cloud-chamber-record}). When traversing
a cloud chamber, particles leave traces which resemble trajectories of material objects,
possibly subject to the laws of classical mechanics. Heisenberg explains
that, in light of the new theory, this interpretation of the tracks cannot be upheld. 

\begin{figure}
\begin{centering}
\includegraphics[angle=90,width=0.9\textwidth]{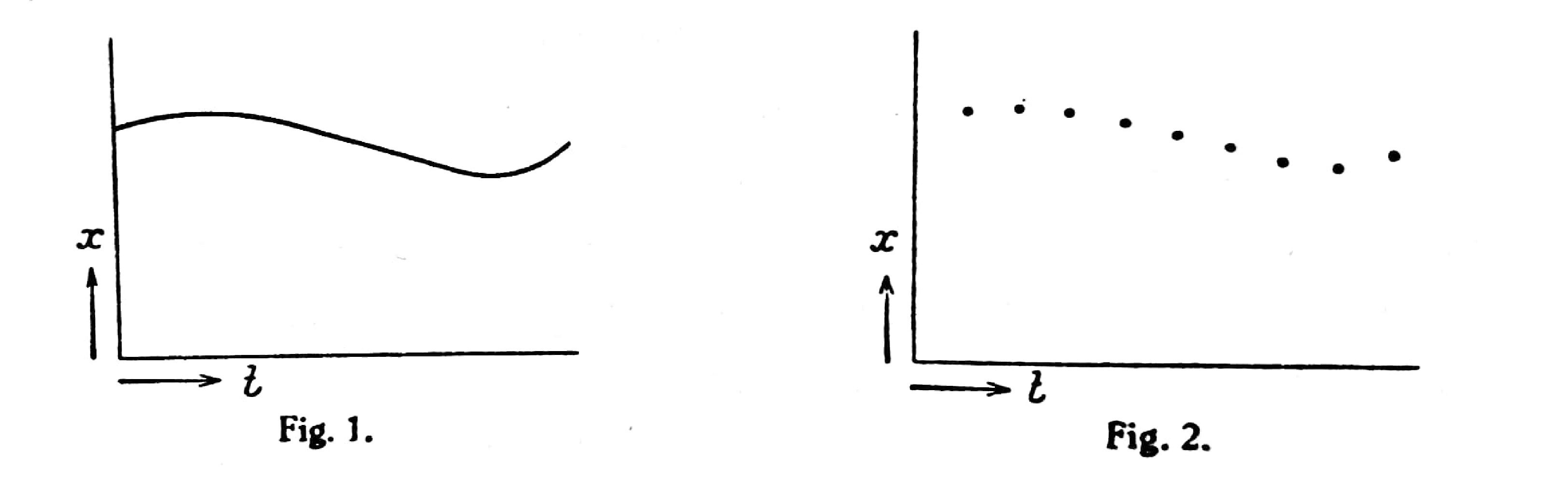}
\par\end{centering}
\caption{\label{fig: cont vs discrete} Smooth trajectory of a classical point particle (left); schematic representation of the track of a quantum particle recorded by a cloud chamber (right). \cite[p 173]{HeisenbergSketch}}
\end{figure}

The classical Newtonian equations of motion for a point particle
predict continuous trajectories by ascribing to it a definite location
and velocity at each moment of time (cf. the sketch on the left in
Fig.~\ref{fig: cont vs discrete}). Heisenberg points out that the
tracks seen in cloud chambers are, in fact, not smooth: careful inspection
of the tracks shows that they rather consist of sequences of condensate droplets which result from the repeated interaction of the particle with the saturated vapor filling the entire chamber (cf. the sketch on the right
in Fig.~\ref{fig: cont vs discrete}).

A sequence of rather imprecise particle positions cannot be used to
reliably determine the velocity of a massive particle. One might hope that refined measurements would resolve this difficulty, a strategy known to be successful when observing large, macroscopic objects such as people or planets. Quantum theory, however, comes with a built-in finite limit on the possibility to
simultaneously attribute exact values of position and velocity to a microscopic article. This property, the content of \emph{Heisenberg's uncertainty
relation}, means that one has to give up the idea that ``it will
be possible to trace the trajectory {[}$\ldots${]}, with the tangent
to the curve indicating the velocity'' \cite{HeisenbergSketch}. 
In other words, an electron simply does not ``possess" values for its position and velocity which could be read off using a suitable device. 

Heisenberg was aware of the fact that a radical departure from classical
concepts was required to formulate an alternative theory. In his 1971 autobiography
he describes the role of cloud chambers photographs when searching for a new conceptual framework:
\begin{quote}
\emph{We had
always said so glibly that the path of the electron through the cloud
chamber existed. But perhaps what we really observed was something
much less. Perhaps we merely saw a series of discrete and ill-defined
spots through which the electron has passed. In fact, all we do see
in the could chamber are individual water droplets which must certainly
be much larger than the electron.} \cite[pp 77-8]{HeisenbergBio} 
\end{quote}

To provide a valid description of electrons and other elementary particles,
quantum theory rejects basic assumptions about particles made by classical mechanics. Without the concepts of particle position and velocity it becomes impossible to write down Newton's equations of motion. Instead of the lacking dynamical variables, the ``wave function'' of a system is introduced to describe its state. The paradigm shift is
completed by postulating Schr\"odinger's equation which governs its evolution in time, replacing Newton's equations. As a major consequence, quantum mechanical predictions are, typically, not deterministic but only probabilistic.

\section{Today's paradigms of physics}

About a century after their inception, quantum theory and the theory
of general relativity (which generalizes the early theory considered
in Sec.~\ref{subsec:Revolution relativity} beyond inertial frames),
continue to define our view of the natural work at a fundamental level.
Over time, the theories have been applied successfully to an ever
increasing range of phenomena without, however, the need to modify
the basic assumptions on which they rest. It seems fair to say that,
taken together, they define a highly successful paradigm which leaves
us with tools to solve puzzles we encounter when studying nature
on both microscopic and astronomical scales. 

Quantum theory has been found to possess considerable explanatory power
regarding the behaviour of molecules, atoms nuclei and their constituents.
Recent developments exploit non-classical properties of quantum systems leading to novel applications and technologies 
such as quantum-cryptographic schemes and efficient ways to process 
information encoded quantum mechanically. General relativity is essential to model the early
universe and to describe its current state. The accuracy of global
positioning systems depends on appropriately taking into account relativistic
effects.

According to Kuhn, the discovery of anomalies which do not fit an
existing paradigm signals the end of a period of normal science. Having summarized
today's paradigm of physics, we are now in a position to approach the main
question of this essay: can we identify anomalies which might indicate a future
revolution upending quantum theory and the theory of relativity? To
do so from within a paradigm is qualitatively different from retrospectively
applying Kuhn's terminology to a paradigm of the past.

It is common practice in a period of normal science to look for theoretical
inconsistencies within the accepted theory or to predict previously unobserved phenomena. An experiment crucial
for the acceptance of the theory of (general) relativity was carried
out by Eddington in 1919. It was confirmed that 
light grazing the sun was bent away from a straight path twice as
much as predicted by a Newtonian approach \cite{EinsteinAllgRT,Eddington1919}.
Later tests of the theory have been carried out with increasing precision
and were positive throughout, culminating in the direct observation
of gravitational waves \cite{GravitWaves} and black holes \cite{blackhols}.
No widely acknowledged anomalies seem to exist which
would suggest the need to alter special or general relativity. Of
course, there are open questions such as the existence of ``dark
matter'' but they do not seem to create full-fledged conflicts with relativity
theory as it stands. 

As for quantum theory, an early apparent inconsistency was described
in a paper by Einstein, Podolsky and Rosen in 1935 \cite{EPR}. Their
thought experiment suggests that quantum theory does not provide a complete description
of the particles such as electrons potentially, hence requiring a fundamental
modification. The ``EPR paradox'' led to long-standing discussions about the foundations
of quantum theory. The paradox was finally resolved experimentally
in favour of the counter-intuitive predictions made by quantum theory
\cite{Aspect: Bell,HensonLoopholeFree}.

A long-standing conceptual problem within quantum theory concerns the ambiguous
status of \emph{measurements} i.e. the interactions of
an experimenter with a quantum system in order to extract information
about its state. If one considers quantum theory as universally valid, then
it should be possible to use the theory to completely describe measurements -- in the end, they correspond to
nothing else that interactions between physical systems.
It is, however, not obvious how to set up such a description. The difficulty can be overcome
by assuming an artificial split of the laboratory into a quantum part
(the observed system) and a classical part (the measuring device).
In spite of its conceptual ambiguity, this approach to measurements is effectively being used whenever experiments are performed with quantum systems, be it to test quantum theory or to apply it to achieve some task. 

The unsatisfactory -- if not anomalous -- status of measurements hangs over the theory ever since it has been noticed
in the early days of quantum theory \cite{SchroedingerQMII}.
Two strategies have been applied to resolve the problem. There are
\textit{interpretations} of the existing theory which claim
to remove the difficulty at the expense of other unpalatable assumptions.
Alternatively, slight \textit{modifications} of the standard formulation of
quantum theory introduce mechanisms which allow one to avoid the problem.
Since these approaches do not lead to observable differences, the
conceptual difficulty simply persists in one way or the other. The measurement problem has been
made quite explicit by refining a paradoxical thought experiment dating
back to the 1960s \cite{Wignersfriend,FRParadox}. Nevertheless, there
is no candidate theory to replace quantum theory in its current form while maintaining or
even extending its predictive power.

Quantum theory and relativity, each on their own, represent successful,
seemingly independent tools to describe the properties of matter on
very small and very large scales, respectively. However, if elementary particles
move at high speeds, both special relativity and quantum theory are
relevant. \emph{Relativistic quantum theory}, or \emph{quantum field
theory}, combine the two into a single consistent framework. Its predictions have been confirmed with remarkable accuracy in particle accelerators. 

In contrast, it is not known how to incorporate the basic assumptions
of quantum theory into the \textit{general} theory of relativity. Einstein
was aware of the rigid, monolithic structure of general relativity,
anticipating the difficulty to modify it in a letter to the London
Times in 1919,
\begin{quote}
\emph{The chief attraction of the theory lies in its logical completeness.
If a single one of the conclusions drawn from it proves wrong, it
must be given up; to modify it without destroying the whole structure
seems to be impossible.} \cite[p 105]{EinsteinTimes1919}
\end{quote}

One of today's major challenges remains to develop a theory which unifies
quantum theory with general relativity. A theory of \emph{quantum
gravity} is needed to describe the early stage of the universe when
extreme densities of matter occur at very small scales, and in the
presence of strong gravitational forces. However, to perform
controlled experiments in such circumstances is unfeasible. In that
sense, physics is not necessarily facing an anomaly as such but rather an
open, admittedly difficult problem within current normal science.

\section{Future revolutions?}

Let us, finally, try to go beyond the original scope of Kuhn's historical approach:
do his concepts also apply to contemporary science, or are they limited to past paradigm
only? Do they allow us to construct a ``meta-scientifc'' vantage point to judge the status of a currently accepted paradigm? Can we identify anomalies which hint at developments potentially overthrowing the theories which determine our current view of the world? 

\begin{wrapfigure}{o}{0.5\columnwidth}%
\centering{}\includegraphics[scale=0.25]{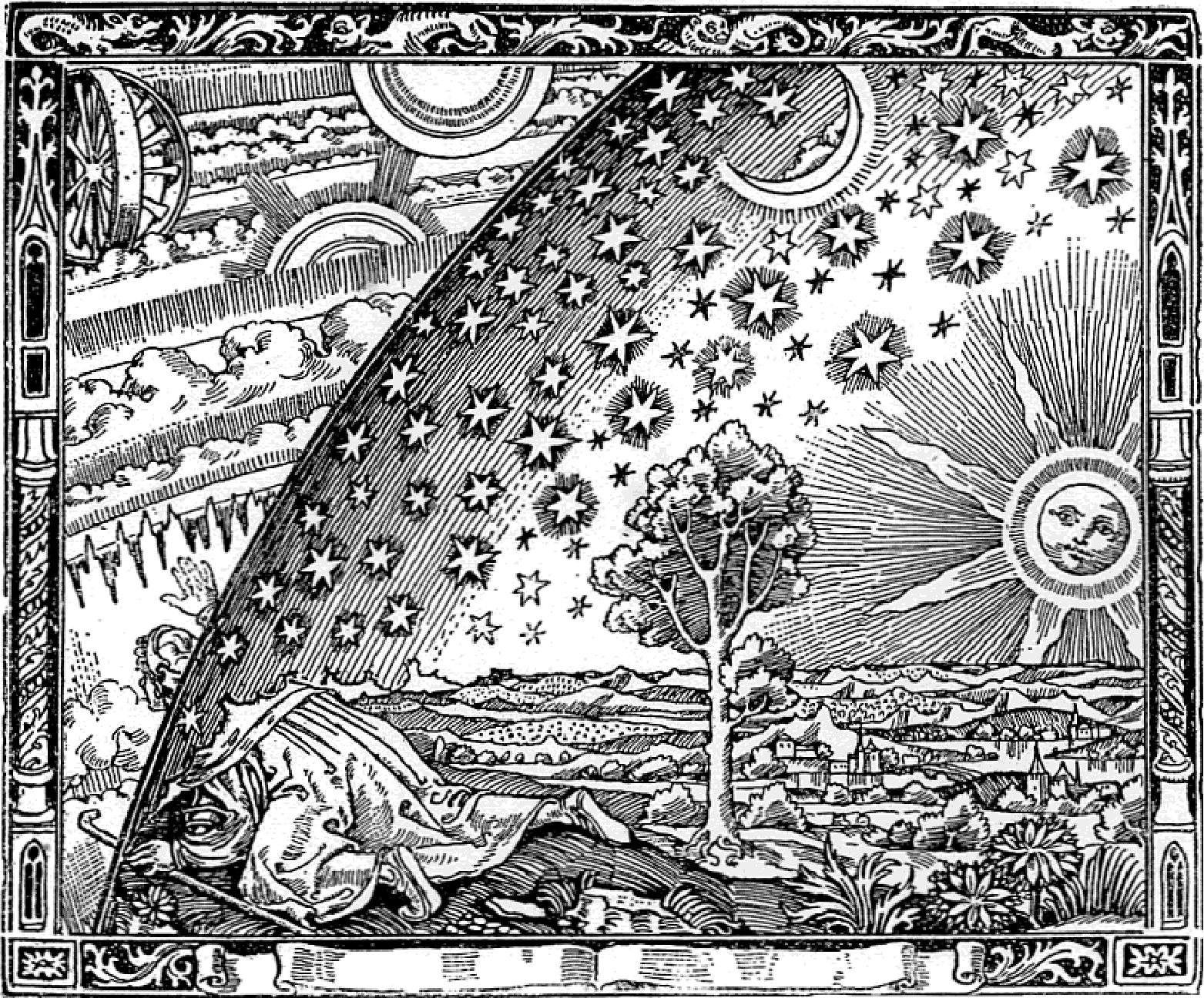}\caption{\label{fig: flammarion engraving} A 19th-century engraving illustrating the solid crystal spheres separating the heavens from the earth \cite[p 163]{flammarion1888}.}
\end{wrapfigure}%

The idea of ``normal science'' clearly applies to today's physics, easily characterized as a paradigm-based undertaking which provides
puzzles and the tools to solve them. Quantum theory has reached a state in which it generates
new technologies, in spite of the long-standing problem of measurement. Thus, the measurement problem may not count as a full-fledged anomaly in Kuhn's sense. The consequences of general relativity are being unfolded in 
astrophysics, by matching an ever larger body of data; solvable puzzles rather than crisis-inducing anomalies are the norm. Thus, taken
by themselves, the theories which form today's well-established framework of physics do
not seem to head towards an imminent crisis. 
 
Historically, the unification of theories has been a strong driving force within physics. In this respect, the current framework of two coexisting
and largely independent theories is unsatisfactory. Flammarion's engraving (see Fig.~\ref{fig: flammarion engraving}) depicts the Copernican revolution:  an inquiring mind has finally broken through the venerable celestial model dating back to antiquity. 
Quantum theory and relativity represent today's crystal spheres which shape our views of the material world. We do not know where to look in order to transcend them. Physicists almost routinely search for inconsistencies of theories by testing their limits. The notion of anomaly does not seem to supply an alternative handle with predictive
power which would make it easier to recognize the start of a revolutionary period.

To conclude, we have seen that Kuhn's proposed structure of scientific revolutions is an
instructive scheme capable to describe how the physical paradigm valid today came into existence. It is not difficult to point to anomalies which triggered revolutionary crises, ultimately causing the removal of Newtonian mechanics, the earlier paradigm. However, Kuhn's concepts appear to remain purely descriptive: they do not provide clues regarding the future development of today's paradigms, the theories of relativity and quantum theory. They continue to cast their spell on us.

\ack This paper is an extended version of a talk given at the one-day HAPP conference ``Paradigm Shifts Across the Ages,'' which took place at the University of Oxford on 8 June 2019. My thanks go to the organiser J Ashbourn of the St Cross Centre for the History and Philosophy of Physics.

\end{document}